# Proposal for experimental verifying of Machian transient mass fluctuations

**Viorel Drafta**

E-mail:  drafta_v@yahoo.com

**Abstract**

Starting from quoted papers it show that, applying Mach's principle, on obtains transient mass fluctuation by varying proper energy.  It is show why the experiments performed until now measured a smaller effect than was predicted. On establish conditions for using gravitational energy by means of transient mass fluctuations. On suggest a new type of experiment for verify the effect of transient mass fluctuation.

## 1. Theoretical overview

It is known  Mach's principle who assert that inertia of the bodies is caused by interaction between bodies and the distant matter from Universe.

Because the gravitation is the only known interaction between different masses, it is natural to suppose that, if  Mach's principle is correct, inertia is an effect due to gravitational interaction [1],[2].

If external force acting on a body accelerate it, as reaction to acceleration appear inertial reaction force acting on external force source. Supposing that above hypothesis is correct, inertial reaction force is due to gravitational interaction between body and the distant matter from Universe. Considering this force as relativistically invariant four-force on divide it by the mass to get the field strength who produce the four-force. Using four dimensional Gauss law (for details see Appendix of [2]), getting:

$$\nabla^2 \Phi - \frac{1}{c^2}\frac{\partial^2 \Phi}{\partial t^2} = 4\pi G \rho_0 + \frac{\Phi}{\rho_0 c^2}\frac{\partial^2 \rho_0}{\partial t^2} - \left(\frac{\Phi}{\rho_0 c^2}\right)^2\left(\frac{\partial \rho_0}{\partial t}\right)^2 - \frac{1}{c^4}\left(\frac{\partial \Phi}{\partial t}\right)^2 \qquad (1)$$

where:

$\Phi$ - total gravitational potential;

$\rho_0$ - rest matter density;

G - constant of gravitation;

c - vacuum light speed.

The strong form of Mach's principle asserts that body's mass is consequence of his gravitational potential energy pertain to the whole matter of Universe. That is:

$$E = m \Phi \qquad (2)$$

As shown by Sciama ([3] quoted in [1]), the gravitational induction of inertial reaction forces requires that the locally measured value of $\Phi$ must be equal to the $c^2$ so (2) have the same meaning with $E = mc^2$.

Equation (1) represents d'Alambertian of potential who is equal with his local sources. On observe that in right part of (1) are four terms. The terms 2 and 3 depends on time variation of mass density (or equivalent, energy density). Those two terms demonstrate that it could obtain mass transient fluctuation of an object by time varying of his energy density.

On observe that third term is always negative, but due to $1/c^4$ factor this term is negligible small. Still, this third term (call it "wormhole" term) could be a source for negative energy densities for Alcubierre type devices.

The fourth term could be considered 0 because time variation of gravitational potential could be neglected.

In this way equation (1) become:

$$\nabla^2 \Phi - \frac{1}{c^2}\frac{\partial^2 \Phi}{\partial t^2} \approx 4\pi G\left(\rho_0 + \frac{\Phi}{4\pi G \rho_0 c^2}\frac{\partial^2 \rho_0}{\partial t^2}\right) \qquad (3)$$

On observe that transient term adjust positively or negatively object's proper mass density. On write:

$$d\rho_0 = \frac{\Phi}{4\pi G \rho_0 c^2} \frac{\partial^2 \rho_0}{\partial t^2} \quad (4)$$

From which on obtain equation for mass fluctuation:

$$dm = \frac{\Phi V}{4\pi G \rho_0 c^2} \frac{\partial^2 \rho_0}{\partial t^2} \quad (5)$$

where V is the volume of body with mass $m_0 = \rho_0 V$

In [1] and [2] are described the experiments for verifying mass transient fluctuations. The results show that appear a force, but smaller than theory predicted.

The authors of experiment assume that the smallness of observed force is due to opposite displacement of positive and negative ions from barium titanate who is subject to electric field. The effect diminished too because a part of energy increase the crystal stress and remaining is recovered as kinetic energy.

The experimenters conclude that the effect exist and future research must find the conditions in which the effect increased.

## 2. Observations and theoretical addendum

### 2.1. Discussions on equation (5)

In equation (5) on can replace body's mass density by body's proper energy, so we can write:

$$dm(t) = \frac{\Phi V}{4\pi G E(t) c^2} \frac{\partial^2 E(t)}{\partial t^2} \quad (6)$$

The above equation show that mass can fluctuate by rapid variation of body's proper energy instead of kinetic energy variation.

For demonstrate this, on calculate the second derivative of body's energy $E=mc^2+\Delta E$ when the body's acceleration is constant (rest mass is $m_0$ and $\Delta E$ is part of energy not due to the mass):

$$\frac{\partial^2 E(t)}{\partial t^2} = m_0 a^2 \frac{1+2\frac{v^2}{c^2}}{\left(1-\frac{v^2}{c^2}\right)^{\frac{5}{2}}} + \frac{\partial^2 \Delta E}{\partial t^2}$$

In classical approximation v<<c or in instantaneous reference frame (v=0), the above equation becomes:

$$\frac{\partial^2 E(t)}{\partial t^2} = m_0 a^2 + \frac{\partial^2 \Delta E}{\partial t^2}$$

Result that mass fluctuation depend on body's acceleration and time variation of $\Delta E$. If $\Delta E$=constant then the below equation shows the dependence of mass fluctuation by body's acceleration:

$\delta m = ka^2$

For constant applied force F on write:

$ka^3 + m_0 a - F = 0$

With solution:

$a = A - m_0/3kA$

Where:

$$A = \left( \frac{F}{2k} + \frac{\sqrt{3}}{18k} \sqrt{\frac{4m^3 + 27kF^2}{k}} \right)^{\frac{1}{3}}$$

Different from:

a=F/m$_0$

This difference was not measured for charges accelerated in strong electric fields, so on can conclude:

a) the variation of kinetic energy do not produce mass fluctuation;
b) electrical charges have a "compensatory" mechanism by which mass don't fluctuate.

The fact that kinetic energy depends on reference frame support conclusion a) above.

In conclusion $\frac{\partial^2 E}{\partial t^2}$ can reduce at $\frac{\partial^2 \Delta E}{\partial t^2}$ and equation (6) on write:

$$\boldsymbol{\delta} m(t) = \frac{\Phi V}{4\pi G E(t) c^2} \frac{\partial^2 \Delta E(t)}{\partial t^2} \tag{7}$$

In this situation the total (relativistic) energy on write:

$$E = \frac{m_0 + \boldsymbol{\delta} m(t)}{\sqrt{1 - \frac{v^2}{c^2}}} c^2 + \Delta E \tag{8}$$

In [1] and [2] on suppose that mass fluctuations are due to kinetic energy variation for accelerated bodies, but equation (7) shows that mass fluctuations are due to proper energy modification whatsoever body is moving. This interpretation lead to many devices for energy production and space transportation.

## 2.2. Addendum for equation (1)

In linearised weak field approximation of General Relativity Theory, by analogy with Electromagnetism, on define gravitational vector potential **A** together with gravitational scalar potential $\Phi$ ([6], [7]):

$$\Phi = G \int_V \frac{r}{r} dV$$

$$\vec{A} = \frac{G}{c^2} \int_V \frac{r\vec{v}}{r} dV$$

If **v**=constant (see discussion in [5]) on write:

$$\vec{A} = \frac{1}{c^2} \Phi \vec{v}$$

Continuing the analogy, the vector *f* from appendix of paper [2], for demonstration of equation (1) must be write:

$$\vec{f} = -\nabla \Phi - \frac{\partial \vec{A}}{\partial t}$$

Who involve following supplemental terms in right member of equation (1):

$$-\frac{1}{c^2}\left[\vec{a}\nabla\Phi + \Phi\nabla\vec{a} + \frac{\partial \Phi}{\partial t}\nabla\vec{v} + \vec{v}\frac{\partial}{\partial t}(\nabla\Phi)\right] \quad (9)$$

For fields with slow time variation the last two terms are negligible, so (9) could reduce to:

$$-\frac{1}{c^2}\left[\vec{a}\nabla\Phi + \Phi\nabla\vec{a}\right] \quad (10)$$

First term from (10) become significant for large gradient of Φ and/or large acceleration. The second term become significant for large div **a**. For strong accelerated particles this effect could be subject of experimental observations.

**2.3. Comments on experiments performed by Prof. J.F. Woodward and Dr. T. Mahood**

The authors of papers [1] and [2] try to include in the same device the propulsion system and the mean by which mass fluctuates. They considered that smallness of observed effect is due to following facts:

- opposite displacement of positive and negative ions from barium titanate who is subject to electric field;
- a part of energy increases the crystal stress and remaining is recovered as helpful kinetic energy.

Because equation (7) shows that mass fluctuations are due to proper energy modification, when apply a pure sinusoidal signal on that device, the analytical expressions of time variation for applied tension, energy in capacitor and second time derivative of that energy are:

$$U(t) = U_0 \sin(\omega t)$$

$$E(t) = E_0 + [CU_0^2/2] \sin^2(\omega t) \qquad (11)$$

$$d^2E(t)/dt^2 = CU_0^2 \omega^2 \cos(2\omega t)$$

Simplifying the crystal model, on suppose a center of positive ions and a center of negative ions, with a minimum distance between them. Applying a sinusoidal electrical tension on produce a relative displacement of those two centers, correlated with the amount of tension. Considering the mass center reference frame, the expressions for time variations of positions, velocities and ion's masses are (notation is obvious):

$$x_+ = x_{0+} \sin(\omega t)$$
$$x_- = -x_{0-} \sin(\omega t)$$

$$v_+ = v_{0+}\omega \sin(\omega t)$$
$$v_- = -v_{0-}\omega \sin(\omega t)$$

$$m_+ = m_{0+} + \delta m_{0+} \cos(2\omega t)$$
$$m_- = m_{0-} + \delta m_{0-} \cos(2\omega t) \tag{12}$$

By replacing in equation for force:

$$F = d(mv)/dt$$

on obtain that force as addition of trigonometric terms with consequence that it cannot appear a net force. The same calculus for an $\nu$ and $2\nu$ mixed signals lead to the same result.

Looking at figure 15 from [1] where are shown the desired and real form of signals applied to the device, on see that real signal is almost sinusoidal and it is much different than desired one. The asymmetry of applied signal generates the net force whose amount depends on asymmetry amount. Also, the inhomogenities of crystal structure lead to the inhomogeneous variation of energy that lead to smallness of effect. I think those are the explanations for diminishing real net force relative to predicted force.

On can withdraw the following conclusions:

- if the device could realize a significant amount of mass modification and force is generate separately, during mass reduction a small amount of force could produce a large acceleration;

- to gain a large mass fluctuation it must vary very fast the body's proper energy and this variation must be very asymmetric. It can be found such a shape that mass only reduce;

- it is desirable to use systems with minimum of inhomogenities such as low temperature systems or with high coherence behavior.

## 3. Optimum shape for energy variation

If body's energy varying periodically according to equation below:

$$E(t) = \begin{cases} E_0 \dfrac{t}{\Delta t} & for \quad t \in (0, \Delta t] \\ E_0 \left[ 1 - \dfrac{(t - \Delta t)^2}{(T - \Delta t)^2} \right] & for \quad t \in (\Delta t, T] \end{cases} \quad with \quad \Delta t < T \quad (13)$$

The second time derivative varies according to equation below:

$$\frac{\partial^2 E(t)}{\partial t^2} = \begin{cases} 0 & for \quad t \in (0, \Delta t] \\ -\dfrac{2E_0}{(T - \Delta t)^2} & for \quad t \in (\Delta t, T] \end{cases} \quad (14)$$

The diagrams for those two quantities are in figures 1 and 2.

On observe that for range (0, Δt] the second time derivative is zero, as well as the mass variation. For range (Δt, T] the same quantities are negative.

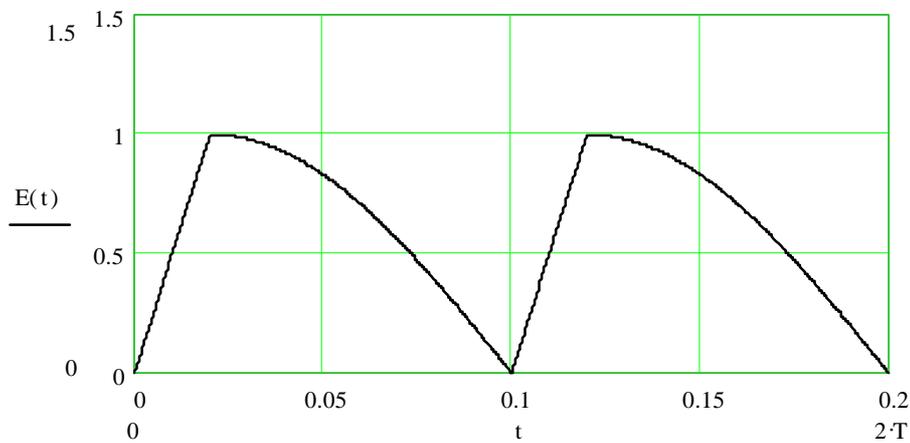

**Fig. 1**

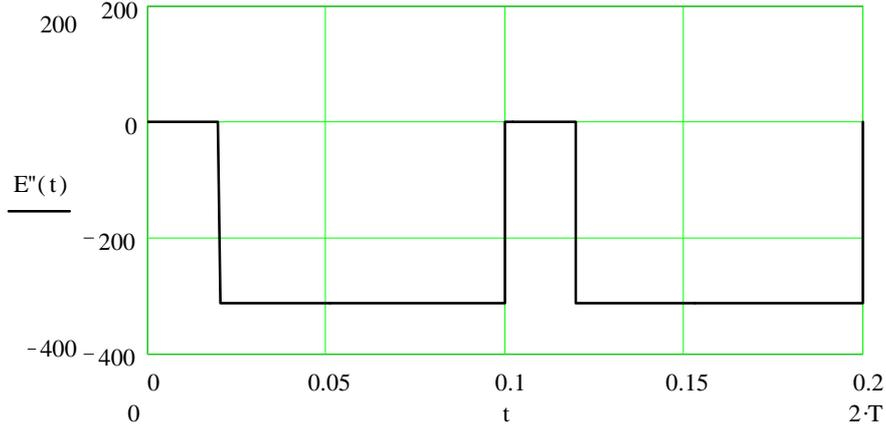

**Fig. 2**

Choosing proper frequency and amplitude for energy variation on can obtain a periodically significant decrease of mass.

## 4. Using gravitational energy by means of transient mass fluctuations

In quoted papers ([1], [2] and [5]) on emphasize that, if Mach's principle is real, the conservation laws for energy, momentum (and angular momentum) are nonlocal. This fact could be used to extract energy from gravitational energy.

Let us take a body with proper mass $m_0$, momentum $\mathbf{p_0}$, angular momentum $\mathbf{L_0}$ and kinetic energy $E_{c0}$ in a given reference frame. On consider the classical approximation $v \ll c$.

For energy extraction on take the following cycle:

1. on decrease body's proper mass by suitable proper energy variation:

$$m = m_0 - \delta m \tag{15}$$

In the same time it produces a momentum, angular momentum and kinetic energy transfer towards distant masses. The new values of those quantities are:

$$\mathbf{p}_2 = m\mathbf{v}_0$$
$$\mathbf{L}_2 = m(\mathbf{r_0} \times \mathbf{v_0}) \tag{16}$$

$E_{c2} = mv_0^2/2$

2. On apply a force who accelerates the body during his mass decreased. At end of acceleration the new values of momentum, angular momentum and kinetic energy are:

$\mathbf{p}_3 = m\mathbf{v}$
$\mathbf{L}_3 = m(\mathbf{r} \times \mathbf{v})$, with $v > v_0$ (17)
$E_{c3} = mv^2/2$

3. The body's mass become normal and it take place a inverse transfer of momentum, angular momentum and kinetic energy:

$\mathbf{p} = m_0\mathbf{v}$
$\mathbf{L} = m_0(\mathbf{r} \times \mathbf{v})$ (18)
$E_c = m_0 v^2/2$

4. On repeat the cycle.

On nominate $\Delta E_f$ the energy consumed for realize mass fluctuation. The energy variation during entire cycle is:

$$\Delta E = (m_0/2 - m)(v^2 - v_0^2) - \Delta E_f \qquad (19)$$

On see that if $m < m_0/2$ and $(m_0/2 - m)(v^2 - v_0^2) > \Delta E_f$ then on could extract useful energy from gravitational energy.

## 5. Proposal for experimental verifying of Machian transient mass fluctuations

Let us take a positive paramagnetic ion with mass $m_0$, charge q and magnetic momentum $\mu$. Let it be the experimental configuration showed in figure 3. The notation represents:

- S - source for positive ions with small velocities v<<c;
- C - device for selecting ions with velocity **v**$_0$;
- a - acceleration sector;
- f - braking sector;
- D - detector;
- **E**$_a$, **E**$_f$ - acceleration and braking electric fields;
- **B** - uniformly magnetic field.

The whole device is in high vacuum.

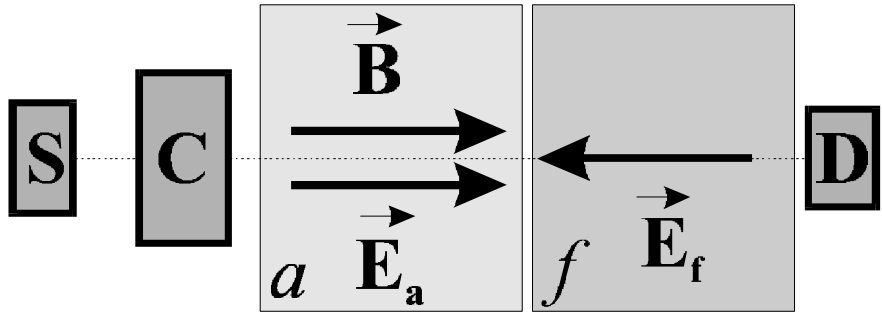

**Fig. 3**

Among ions from source are selected those who have velocity near to **v**$_0$. The selected ions get into acceleration sector. Because they have almost the same velocity, their relative velocities are small and equivalent temperature is few °K. In those conditions almost all ions magnetic momentum becomes oriented parallel to magnetic field. Their energy $U_0 = -\mu B$ is lower. During time T that ions passing through accelerating sector, the magnetic field induction is modified from $B_0$ up to $B_0+\Delta B$ pursuant to equation below:

$$B = B_0 + t^2/T^2 \Delta B \qquad (20)$$

In this case the ion's proper energy decrease from $U_0= -\mu B_0$ down to $U = U_0-\mu\Delta B$ by a relationship corresponding to (13). The second time derivative of ion's energy is negative and ion's mass decrease during T.

In this case the ions gain more kinetic energy than their mass remain at initial value.

If on choose braking field strong enough, the ions cannot get to detector if their mass not decreased during acceleration phase, but they reach the detector if their mass decreased during the same phase.

**Note:** for the above experiment it can be used electrons, but for the same effect T must be shorter with several orders of magnitude.

## 6. Conclusions

The experiment described above allow to verify in a convenient way and without doubt if Machian transient mass fluctuations are real.

If experiment will succeed, this opens the possibility to realize new kind of devices for energy production and space transportation.

**Acknowledgments** to Dr. Martin Tajmar who brings my attention on paper [1].

## 7. Bibliography

1. J.F. Woodward, T. Mahood, "What is the Cause of Inertia?", *Foundations of Physics*, Vol. 29, No 6, 1999
   or
   J.F. Woodward, T. Mahood, "Mach's Principle, Mass Fluctuations and Rapid Space-time Transport", http://chaos.fullerton.edu/Woodward.html

2. T. Mahood, "A Torsion Pendulum Investigation of Transient Machian Effects", http://www.serve.com/mahood/thesis.pdf

3. Sciama, D., 1953, "On the Origin of Inertia," *Monthly Notices of the Royal Astronomical Society* **113**, 34-42.

4. J.F. Woodward, US Patent No. 5280864

5. J.F. Woodward, "The Origin of Inertia", "Radiation Reaction", "Transient Mass Fluctuations", "Killing Time", http://chaos.fullerton.edu/Woodward.html

6. M. Tajmar, C.J. de Matos, "Coupling of Electromagnetism and Gravitation in the Weak Field Approximation", http://xxx.lanl.gov/gr-qc/0003011

7. C.J. de Matos, M. Tajmar, "Advance of Mercury Perihelion Explained by Cogravity", http://xxx.lanl.gov/gr-qc/0005040